# Subjective Difficulty in a Verbal Recognition-based Memory Task: Exploring Brain-behaviour Relationships at the Individual Level in Healthy Young Adults

Jason Steffener[1*], Chris Habeck[2], Dylan Franklin[3], Meghan Lau[1], Yara Yakoub[1], Maryse Gad[1]

1 Interdisciplinary School of Health Science, University of Ottawa, Ottawa, ON Canada

2 Cognitive Neuroscience Division, Department of Neurology and Taub Institute for Research on Alzheimer's Disease and The Aging Brain, Columbia University College of Physicians and Surgeons, New York, New York, United States of America

3 School of Psychology, University of Ottawa, Ottawa, ON, Canada

* Corresponding Author

Jason Steffener, PhD

Interdisciplinary School of Health Sciences

University of Ottawa,

200 Lees, Lees Campus,

Office # E250E,

Ottawa, Ontario, CANADA

K1S 5S9

Email: jsteffen@uottawa.ca

**Keywords:** brain-behavior, subjective difficulty, verbal recognition-based memory, fMRI, individual level mapping, sex differences

**Running Title:** Brain-behaviour Relationships at the Individual Level

**Summary:** 6806 Words, 9 Figures, 2 Tables

## Highlights

- Task-related responses to brain activity mapped at the individual level.
- Patterns of brain activity differ across individuals.
- The utilization of individual maps of brain activity is similar across a group.
- The use of individual maps is a better predictor of behavior than group-derived maps.
- Expression of the group and individual patterns differed between the sexes.

## Data Availability

The data that support the findings of this study are openly available in Center for Open Science at https://osf.io/w9vte/

## Conflict of Interest

The authors have no conflicts of interest to declare.

## Abstract

The vast majority of fMRI studies of task-related brain activity utilize common levels of task demands and analyses that rely on the central tendencies of the data. This approach does not take into account perceived difficulty nor regional variations in brain activity between people. The results are findings of brain-behavior relationships that weaken as sample sizes increase. Participants of the current study included twenty-six healthy young adults evenly split between the sexes. The current work utilizes five parametrically modulated levels of memory load centered around each individual's predetermined working memory cognitive capacity. Principal components analyses (PCA) identified the group-level central tendency of the data. After removing the group effect from the data, PCA identified individual-level patterns of brain activity across the five levels of task demands. Expression of the group effect significantly differed between the sexes across all load levels. Expression of the individual level patterns demonstrated a significant load by sex interaction. Furthermore, expressions of the individual maps make better predictors of response time behavior than group-derived maps.

# Introduction

The majority of functional brain imaging studies rely on the interpretation of group analyses. Group analyses identify the brain regions where the central tendency of activity is common to the participants of each group within a study. This approach defines individual differences as noise or measurement error around the central tendency. It is becoming recognized that a greater understanding of the between-participant variance is needed to fully understand the functioning brain (Lebreton et al., 2019). Methods that explore the range, source, and effect of individual differences provide many future directions for brain imaging. It facilitates the testing for factors that impact cognitive strategy selection, how demographic differences and lifetime exposures impact patterns of brain activity, structure-function relationships, and the links between brain activity and behavior (Seghier & Price, 2018). Furthermore, identifying and understanding the range of normal variations in individual patterns of brain activity provides insight into the changes that occur in neurodegenerative and psychiatric illnesses (Franzmeier et al., 2020; Tik et al., 2021).

Group analyses identify brain regions used similarly across individuals. Brain-behavior assessments then relate behavioral variables, e.g., task performance, against the variance around the identified central tendency. The weakness in this approach is quantitatively confirmed with the observation that as the sample size increases, the strength of brain-behavior relationships decreases (Grady et al., 2021). These authors explain this phenomenon with the idea that individuals each utilize unique patterns of brain activity when performing the same task. Therefore, the larger the sample, the larger the number of individual patterns of brain activity for performing a task.

Another explanation for individual variance in brain activity when performing a task is that individuals may experience the same task demands differently (Lebreton et al., 2019). These authors argue that tasks are often developed to produce robust population effects and not inter-individual differences. The use of overarching decisions about task demands for all participants in a sample adds variance to the data. This experimentally induced variability results from individuals each experiencing a task with differing levels of subjective difficulty. The various levels of perceived task difficulty likely result in differential levels of brain activity responses. The authors suggest that range adaptation coding principles overcome the limitations of using the same task for all participants. Range adaptation provides each individual with similar subjective input in the form of matched difficulty.

Another source of individual variance in brain activity is sex and gender. A recent review of structural and functional brain imaging results concluded that the brain is not sexually dimorphic (Eliot et al., 2021). However, they concluded that if sex differences did exist, they are buried within individual differences and not detected with group analyses. This conclusion bolsters Grady et al.'s (2021) conclusion that individuals utilize unique patterns of brain activity when performing the same task, limiting the sensitivity of group-level analyses (Grady et al., 2021). Therefore, sex differences in brain activity may not be in the strength of activation within a common brain region but in how an individual performs a task and the regions used.

The current work drops two of the main assumptions of group studies. It does not identify a central tendency in areas of brain activity responding to task demands across all individuals in a group. Instead, individualized patterns of brain activity across multiple levels of task demand are identified. Group level comparisons then utilize these participant-specific patterns of brain activity across task demands. Secondly, tasks are administered based on subjective difficulty instead of global decisions like the number of items to remember. Finally, sex is included in all models to identify whether sex differences exist in how an individual's brain activity responds to task demands.

Multivariate analyses identified the individualized patterns of brain activity (Moeller et al., 1996; Spetsieris & Eidelberg, 2010) while participants were engaged in a delayed match to sample working memory task (Rypma et al., 1999; S. Sternberg, 1966). Multivariate analyses utilize the entire brain in their analyses, unlike univariate analyses, which focus on one location at a time. Focus on brain-wide patterns, therefore, identifies subtle and consistent differences in brain function and minimizes false positives by minimizing the number of statistical tests. Such analyses allow the identification of patterns of brain activity that are specific to each individual. The expression, or utilization, of these brain activity patterns, is then related to a range of cognitive demands. The range of cognitive demand used during fMRI scanning is determined based on each individual's working memory capacity. Demands are therefore parametrically manipulated at levels such as 75, 100, 125% of an individual's cognitive capacity.

This work addresses the research question of whether, within a sample of twenty-six young adults, differential patterns of brain activity are utilized in similar or different manners to meet increasing task demands and whether there are differences in utilization between the sexes. The null hypothesis is that individuals use a common set of brain regions similarly as task demands increase for both sexes resulting in similar performance. In addition, this work tests whether an individuals' utilization of participant-specific patterns of brain activity is a better predictor of cognitive performance than an individuals' utilization of a group-derived pattern of brain activity.

## Methods

### Participants

Participants for this study were recruited from the University of Ottawa. Inclusion criteria required participants to be between 18 - 30 years old, right-handed, normal or corrected to normal vision, and English as a first language. In addition, eligible participants must be in good self-reported health, have no past incidence of a severe head injury, and not be severely ill or hospitalized within the past six months. Participants were excluded if any neurological disorders were likely to affect cognitive function, the use of any psychoactive drugs, and significant cardiovascular disease or atherosclerosis. The study received ethical approval from the Research Ethics Board (REB) of the University of Ottawa, and all participants signed informed consent forms.

**Behavioral Task**

The delayed match to sample task used visually presented letter stimuli (Rypma et al., 1999; S. Sternberg, 1966) with relatively minor adaptations from previous use with the task (Hillary et al., 2003; Steffener et al., 2009; Stern et al., 2008). This experiment is trial-based, where each trial consists of three parts, see Figure 1. A trial begins with the presentation of letters in a 3x3 grid on the screen for 2.5 seconds. The participant is to study and remember the letters. After removing letters from the screen, a green crosshair appears for 3.5 seconds. During this time, the participant is to remember the studied letters. Finally, a single probe letter appears for 2.5 seconds in the center of the screen. During this time window, participants are to determine whether or not they recognize the probe letter as one they studied for this trial. Responses were recorded via a keyboard press.

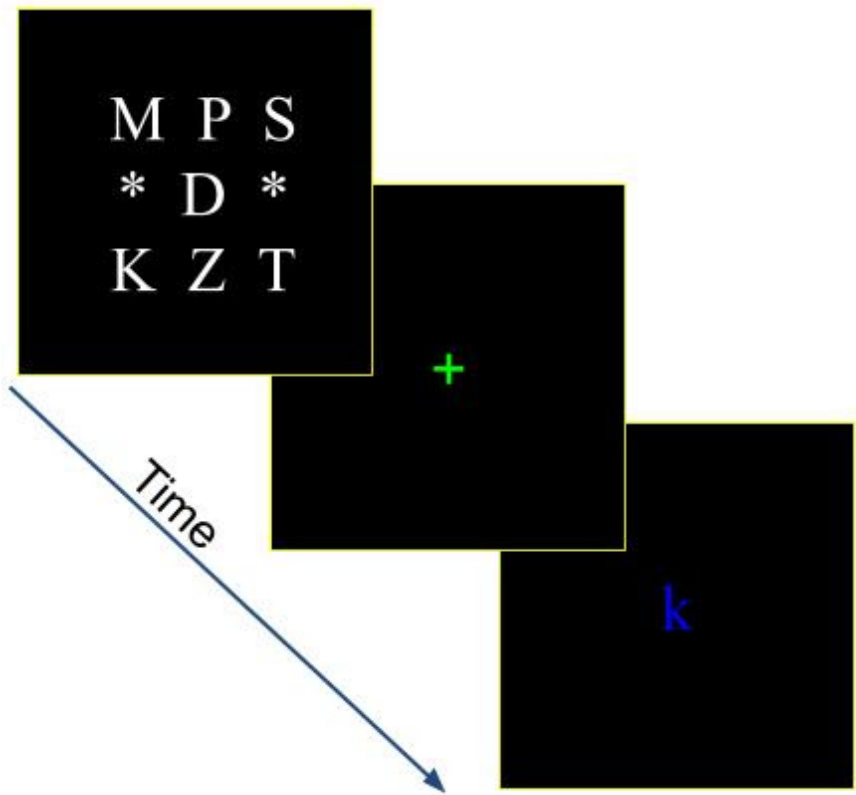

Figure 1
A single trial of the simultaneous presentation of verbal stimuli in the delayed match to sample experiment. The three panels represent the three phases of the experiment. The stimulus encoding phase presents one to nine letters in white font color; the rehearsal phase presents only a crosshair. The recognition phase presents a single probe letter in blue font color. [COLOR IS REQUIRED]

The memory demands for this task were manipulated based on the number of letters to study at each trial with nine levels of demand possible. When the load level for a trial is less than nine, asterixis fill the remaining locations in the grid. The screen position of stimulus letters was the same for all trials of a particular load level. As an example, in Figure 1, a trial with seven letters is presented. For all trials with seven letters, the stimulus letters will always be in these positions. The use of only consonant letters (except "W") minimized word-forming in the studied letters. No letters in one trial can contain any of the letters used in the previous trial. Letters are not presented in alphabetical order, and study letters are presented in uppercase while probe letters are in lowercase. The use of a lowercase probe ensures that letters are not being matched based on visual features. The task was implemented and administered using PsychoPy2 (Peirce et al., 2019). All software to deliver this task is publicly available at https://github.com/NCMlab/CognitiveTasks.

**Cognitive Capacity**

Cognitive capacity is the level of task demand at which a participant consistently performed at 80% accuracy assessed using an adaptive difficulty version of the task, see Figure 2. The adaptive difficulty used a three-up, one-down staircase design. The procedure started with the lowest level of task demand, one letter, and initially increased in one letter of difficulty for each correct response. After making the first incorrect response, task demands decreased by one letter. After the initial period, task demands increased by one letter after three correct responses in a row. This three up/one down staircase procedure provides approximately 80% accuracy. Each point where the direction of difficulty changed was considered a reversal. Cognitive capacity is the average task demands across all reversal points. This procedure was limited to 7 minutes or 20 reversals (Karmali et al., 2016).

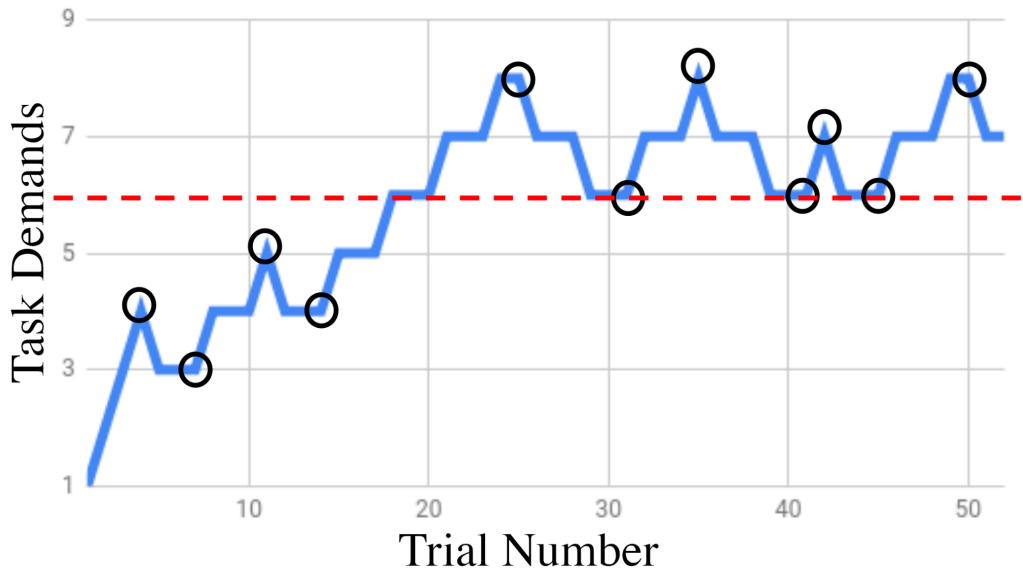

Figure 2 A plot of one individual's changing task demand. As the participant made correct and incorrect responses, the number of letters presented (the task demands) changed. After the initial increase in task demands, after three responses in a row, task demands increased. After one error, task demands decreased. Changes in direction are reversals, as seen circled above. Someone's cognitive capacity is the average of all task demands at reversal values; this is the dashed line for this example. [COLOR IS NOT REQUIRED]

**MRI Data Collection Parameters**

All neuroimaging used the 3T Siemens Biograph mMR MR-PET scanner at the Brain Imaging Centre (BIC) at the Royal Ottawa Mental Health Centre (ROMHC). Participants wore protective earplugs during the scans and held a squeeze ball they could activate if they felt uncomfortable and wished to terminate the scan.

**Structural MRI**

A T1-weighted multi-echo magnetization prepared rapid acquisition gradient echo (MEMPRAGE) image was acquired sagittally (TR = 2530ms; TE 1/2/3/4 = 1.69/3.55/5.41/7.27ms; flip angle = 7°; 1mm isotropic resolution; 192 slices; 256mm field of view, ipat (acceleration) = 2, with 32 ref lines and a non-selective inversion time of 1150 ms and 650 Hz/Px BW). Duration: 6:03 minutes (van der Kouwe et al., 2008).

**Functional MRI**

The following acquisition parameters were used for all task-based data collection. A multi-band accelerated EPI sequence (Moeller et al., 2010) using an acceleration factor of 6, TR =

1110ms, TE = 16.6ms, 52-degree flip angle, phase partial Fourier 6/8, 56 slices collected in an alternating increasing slice order, 2.5x2.5mm in-plane resolution, slice thickness = 2.75mm, field of view: 220x200mm. Each run is 6:38 minutes long.

**Pre-processing**

All image pre-processing and statistical analyses used SPM12 (Wellcome Department of Cognitive Neurology). For each participant's EPI dataset, images were temporally shifted to correct for slice acquisition order using the middle slice acquired in the TR as the reference. All EPI images were corrected for motion by realigning to the first volume of the first session. The T1-weighted (structural) image was coregistered to the first EPI volume using mutual information. This coregistered high-resolution image was used to determine the transformation into a standard space defined by the Montreal Neurological Institute (MNI) template brain supplied with SPM12 using the new segment tool. This transformation was applied to the EPI data and re-sliced using 4th degree B-spline interpolation to 2 x 2 x 2 mm. Finally, all images were spatially smoothed with an 8 mm FWHM kernel.

**Participant level time-series analysis**

The time series modeling for each participant had five regressors of interest, one for each level of memory load. Each block was modeled as a rectangular epoch of 56 seconds in duration. All regressors of the time series models were convolved with a standard double-Gamma model of the hemodynamic response function (Gary H. Glover, 2011; G. H. Glover, 1999). Masking was explicitly applied using all spatially normalized voxels identified as belonging to the brain. The two sessions were modeled together at the first level statistical modeling phase and combined via contrasts. Five contrasts modeled each level of task demand across the two sessions.

**Multivariate Analyses**

Participant-specific patterns of brain activity were identified using scaled subprofile modeling on participant-level data after removing the group-level effect. The result is an individualized pattern of brain activity independent of group effects and the expression of this pattern across all levels of task demand.

For each participant, the five contrast images, one for each level of task demand, were analyzed using scaled subprofile modeling within each participant separately. Scaled subprofile modeling (Moeller et al., 1987; Spetsieris & Eidelberg, 2010) identified the principal components of brain activity across the five levels of task demand. This analysis used the Generalized Covariance Analysis toolbox ([https://www.nitrc.org/projects/gcva_pca](https://www.nitrc.org/projects/gcva_pca)) (Habeck et al., 2005; Habeck & Stern, 2007). This approach produced a series of principal component images and their respective subject scaling factors (SSF), which are each individual's expression of the respective principal component. Only the first principal component for each participant was retained.

The stability of each voxel was assessed using 5000 bootstrap resamples and tested with the percentile method for calculating confidence intervals. Regions were identified as being significantly active if they had a Z score magnitude > 2.

**Statistical Analyses**

Mixed-level modeling tested for load and sex-related effects in task accuracy and response time behavioral measures while controlling cognitive capacity. Mixed models also tested for load and sex effects in brain imaging measures of pattern expression while controlling for cognitive capacity. Mixed level modeling is similar to repeated measures ANOVA except that it has the additional benefit of allowing each participant to have their random intercept and is superior at controlling type I errors  (Barr et al., 2013; Judd et al., 2012). The intercept was a random effect across participants, while load, sex, and cognitive capacity were fixed effects. Model estimation used restricted maximum likelihood, and degrees of freedom were estimated using the Satterthwaite method (Satterthwaite, 1946). Testing for the significance of the random effect used the likelihood ratio test. A significant result demonstrates significant variability in intercept values across participants. The interclass correlation (ICC) value is reported, which is the proportion of the total variance in the dependent variable that is accounted for by the random intercept of each participant (Nakagawa & Schielzeth, 2010). It is the proportion of variation in the data attributed to between-participant differences. In the context of identifying cross-participant similarities, i.e., group effects, the smaller value, the better. A value of zero means that the simpler repeated-measures ANOVA would be as appropriate as the more complex mixed-level modeling. Analyses used Jamovi 1.6.23.0 (Jamovi team, 2019; Jonathon Love, 2019; R Core Team, 2018).

# Results

The following sections summarize the results regarding the participants, the behavioral data, the imaging results, the load and sex relationships with brain imaging data, and the brain-behavioral relationships.

**Participants**

Data from a total of twenty-six participants were included in this study. The mean (standard deviation) age was 22.9 (2.71), with a range of 19 - 30 years. Self-reported sex was thirteen female and thirteen male, all reporting as cisgender.

**Behavioral Analyses**

The mean (std) cognitive capacity was 7.19 (1.06)  letters ranging from 4.62 to 8.44. The accuracy and response times across load levels are shown in Table 1. Cognitive capacity did not significantly differ between sexes ($t$(24) = -0.423, p = 0.676, effect size (Cohens' d) = -0.166; mean (std) females = 7.10 (0.97), males = 7.28 (1.18)). To summarize the following results, there was only a significant main effect of load when predicting accuracy. For response time, the main effects of load and cognitive capacity were significant. Details are described below.

Table 1. Behavioral data across load level

|  | Accuracy Mean (std) | Response Time Mean (std) |
|---|---|---|
| Relative Load 1 | 0.95 (0.092) | 0.75 (0.22) |
| Relative Load 2 | 0.99 (0.033) | 0.82 (0.30) |
| Relative Load 3 | 0.91 (0.11) | 0.97 (0.36) |
| Relative Load 4 (Cognitive Capacity) | 0.83 (0.16) | 1.07 (0.38) |
| Relative Load 5 | 0.78 (0.12) | 1.08 (0.33) |

**Accuracy**

Predicting accuracy, the main effect of load was significant ($F_{(4, 96)} = 16.411$, $p < 0.001$). The interaction was not significant: load by sex ($F_{(4, 96)} = 0.571$, $p = 0.684$). The remaining main effects were also not significant: cognitive capacity ($F_{(1,23)} = 0.872$, $p = 0.360$), sex ($F_{(1,23)} = 0.043$, $p = 0.837$). Repeated differences between levels of task load demonstrate significant differences in accuracy between loads 2 to 3 ($t_{(92)} = 2.686$, $p = 0.009$), and loads 3 to 4 ($t_{(92)} = 2.685$, $p = 0.009$). The remaining differences were not significant: loads 1 to 2 ($t_{(92)} = -1.447$, $p = 0.151$) and loads 4 to 5 ($t_{(92)} = 1.647$, $p = 0.103$). The random component of the model (participant, intercept) was not significant (ICC = 0.012, $X^2(1) = 0.032$ $p = 0.859$).

**Response Time**

Predicting response time, the main effect of load was significant ($F_{(4, 96)} = 20.870$, $p < 0.0001$) as was the main effect of cognitive capacity ($F_{(1, 23)} = 7.097$, $p = 0.0139$). The interaction was not significant: load by sex ($F_{(4, 96)} = 0.335$, $p = 0.854$). The main effect of sex was not significant ($F_{(1,23)} = 3.594$, $p = 0.071$). Repeated differences between levels of task load demonstrate significant differences in response time between loads 2 to 3 ($t_{(92)} = -3.277$, $p = 0.0015$), and loads 3 to 4 ($t_{(92)} = -2.051$, $p = 0.043$). The remaining differences were not significantly different: loads 1 to 2 ($t_{(92)} = -1.619$, $p = 0.109$) and loads 4 to 5 ($t_{(92)} = -0.323$, $p = 0.748$). The random component of the model (participant, intercept) was significant (ICC = 0.673, $X^2(1) = 74.674$, $p < 0.0001$).

## Principal Component Analysis of Brain Imaging Data

### Group Effects

Results from applying principal component analysis to all imaging data collapsed across participants, and load produced a pattern of brain activity representing the activity common to all participants. Out of the 130 images used to derive the group pattern, it accounted for 12.48% of the variance. The group pattern accounted for a range between approximately zero and 63% of the voxel-wise variance in each image. These values are plotted in Figure 3. The mean-variance accounted for by the group effect pattern across participants within each load was: 5.19, 12.30, 21.81, 25.56, and 22.73%.

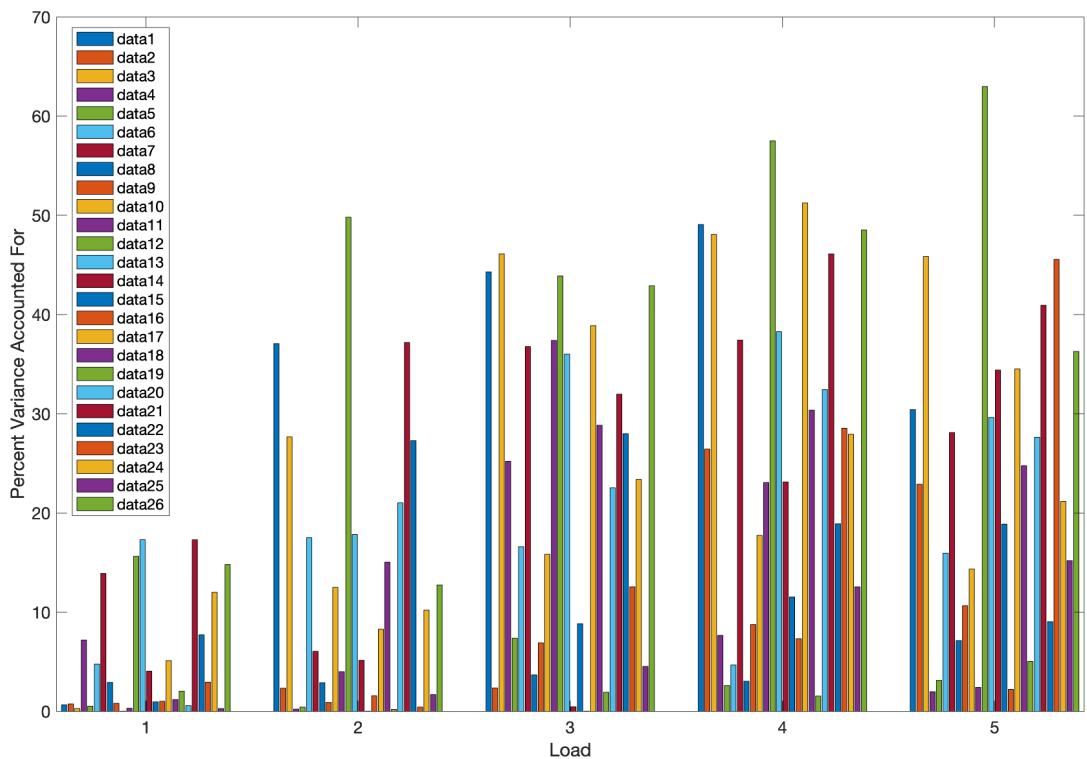

Figure 3. The amount of variance that the group derived pattern of brain activity accounted for in each load-related contrast image from each participant in the study. [COLOR IS REQUIRED]

The group-level pattern of brain activity is shown in Figure 4. A cluster-wise summary of results appears in Table 2. Regions included in the positive direction are the bilateral cerebellum, bilateral superior parietal, medial frontal, and right prefrontal. The following regions are identified in the negative direction: medial frontal, orbital frontal, bilateral angular gyrus, and the precuneus.

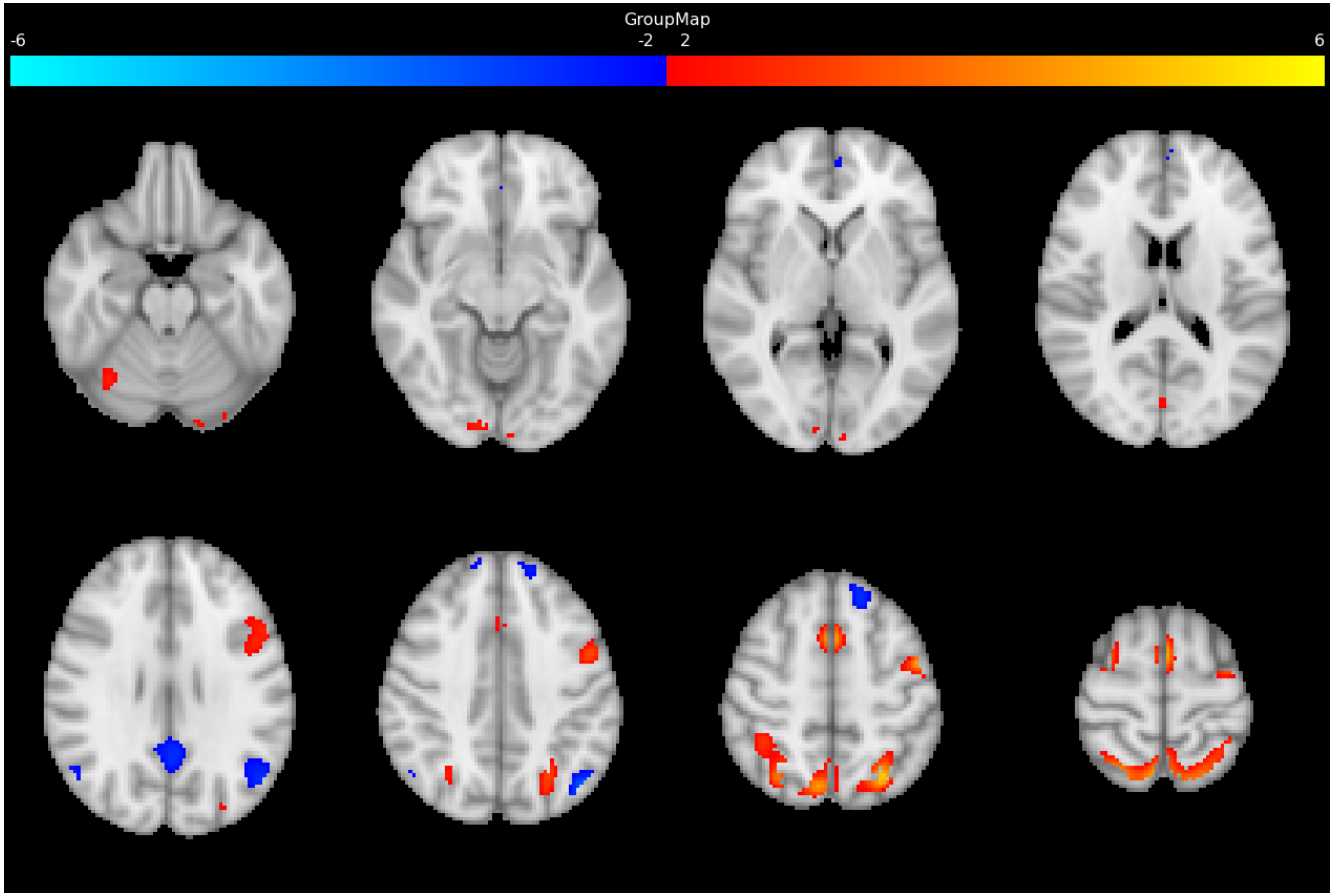

Figure 4. The significance map of the principal component derived from the group data collapsing across participant and load levels. Significance assessed at |Z| > 2.00 using 5,000 bootstrap resamples and the percentile method. [COLOR IS REQUIRED]

Table 2 Group derived pattern of imaging results

| Region | Lat | Xmm | Ymm | Zmm | Z | k |
|---|---|---|---|---|---|---|
| Cerebellum, Crus 1 | L | -30 | -82 | -18 | 2.59 | 286 |
| Cerebellum, 6 | R | 32 | -66 | -20 | 2.80 | 81 |
| -- empty -- | R | 14 | -92 | -16 | 3.45 | 108 |
| Calcarine | L | 0 | -78 | 14 | 2.14 | 20 |
| Precentral | L | -46 | -2 | 52 | 4.46 | 637 |
| Superior Parietal | R | 26 | -62 | 62 | 5.02 | 821 |
| Superior Parietal | L | -26 | -66 | 56 | 6.10 | 876 |
| Supplementary Motor Area | R | 4 | 10 | 52 | 4.09 | 153 |
| Supplementary Motor Area | L | -2 | 4 | 62 | 4.73 | 260 |
| Superior Frontal | R | 28 | 2 | 64 | 3.00 | 68 |
| Medial Orbital Frontal | R | 2 | 50 | -6 | -2.50 | 112 |
| Supramarginal | R | 56 | -26 | 20 | -2.30 | 22 |
| Angular | L | -44 | -68 | 40 | -3.30 | 378 |
| Precuneus | L | 0 | -56 | 32 | -3.01 | 494 |
| Angular | R | 52 | -62 | 30 | -2.48 | 51 |
| Superior Frontal | L | -14 | 30 | 58 | -3.37 | 334 |

Note: Lat: Laterality, mid:midline, k: cluster size.Thresholds were |Z| > 2.00.

**Participant Level Effects**

The group pattern was removed from each of the load-related contrast images from each participant. The PCA was then applied to each participant's five residualized images. From the five contrast images for each participant, the PCA calculated four eigenimages and four eigenvalues. Analyses focused on the first eigenimage; therefore, each participant has their own pattern. To summarize the variance between participants in their maps, Figure 5 shows a voxel-wise map of counts. This map colors voxels based on the number of participants that significantly expressed that location in their pattern. The range in counts was from zero to seven. Therefore, the highest degree of commonality in any brain region was seven individuals. These locations were within the medial prefrontal, supplementary motor areas, posterior parietal, and cerebellum.

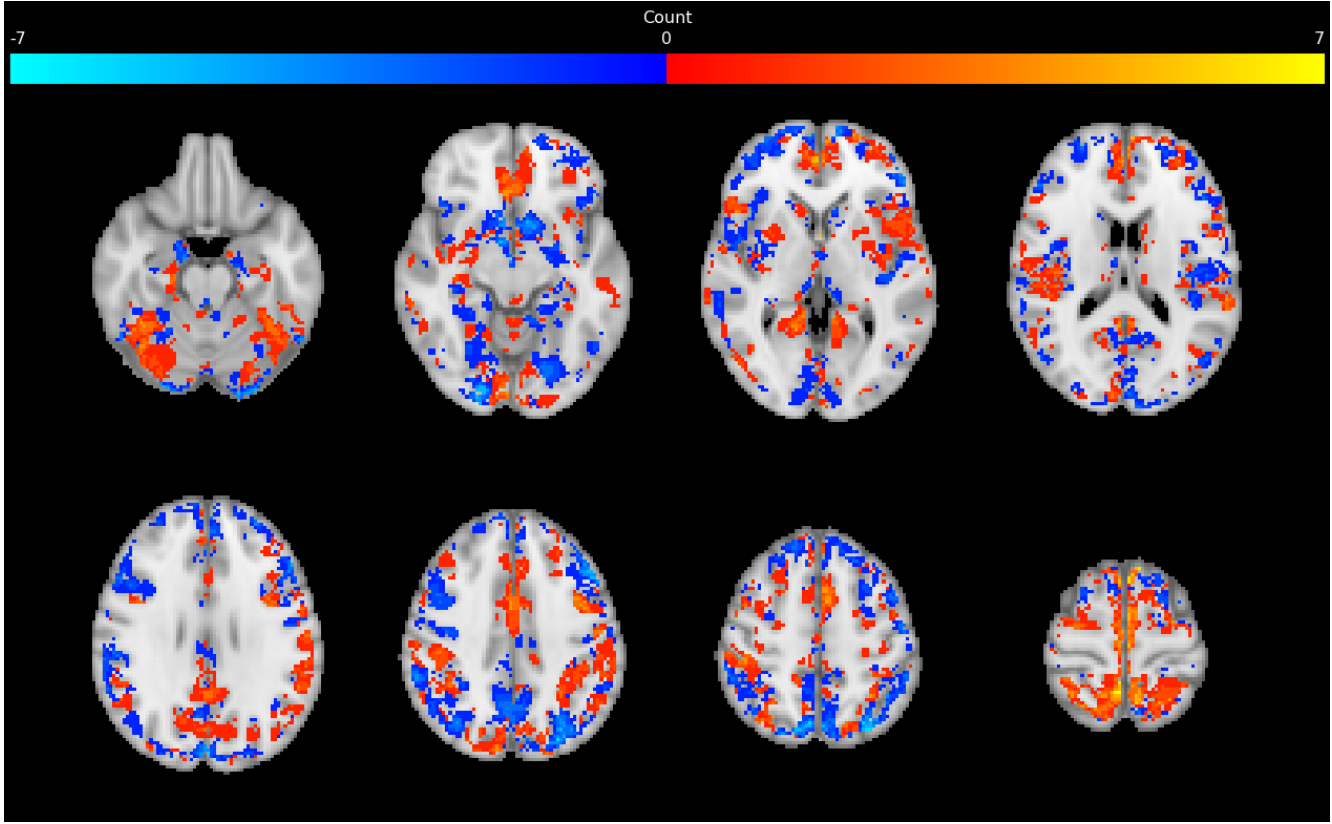

Figure 5. Map of the number of participants whose individual level principal component was significant at each voxel in the two directions. This map is thresholded so that each colored voxel was significant for at least one participant. It is worth pointing out that removing the group effect from the data does not eliminate that voxel from being involved in a participant-level map. Each participant may utilize a region a significant amount above or below the group level. The largest number of participants having significant activation in a voxel is seven, highlighting the large amounts of variability in the spatial distribution of participant-level brain activation. [COLOR IS REQUIRED]

The variance across contrast load images within a participant, accounted for by the first eigenimage, had a mean (std) across participants of 50.26% (8.82%) and a range of 38.59% to 76.63%, see Figure 6. There was no significant relationship between the concentration of variance in eigenimage one and cognitive capacity (r = 0.330, p = 0.100). The amount of variance accounted for by the first eigenimage did not differ significantly between the sexes ($t$(24) = 1.676, p = 0.107, effect size (Cohens' d) = 0.657; mean (std) females = 53% (11.2%), males = 47.4% (4.5%)).

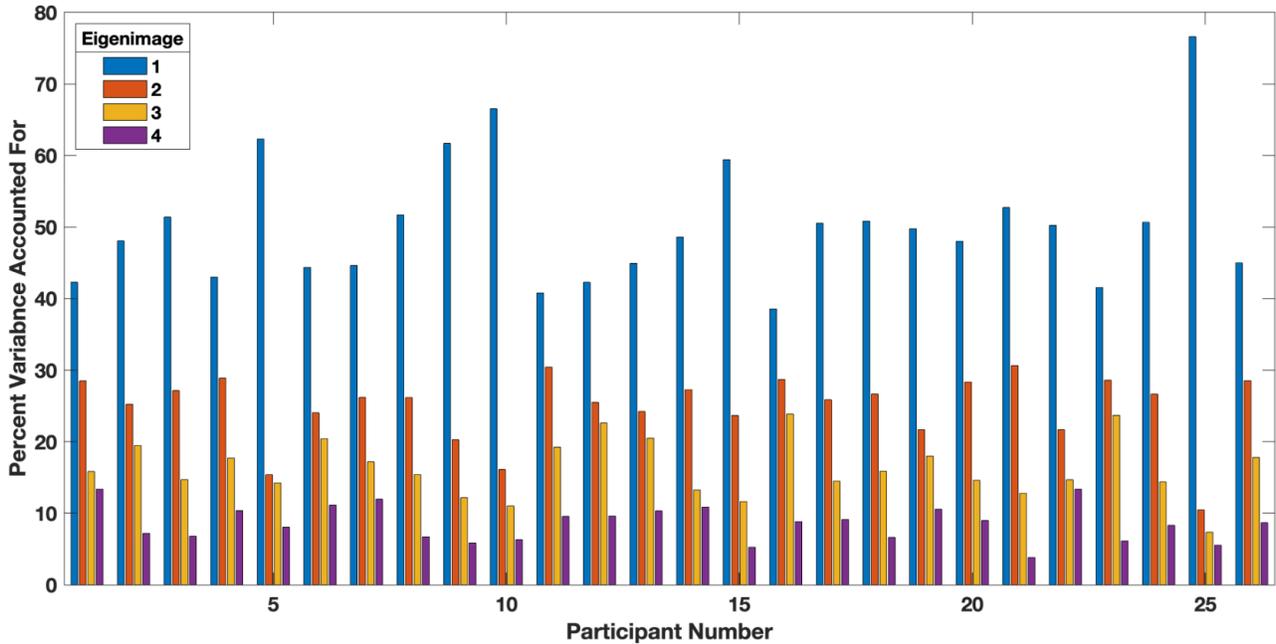

Figure 6. Plots of the percent variance accounted for by the four eigenimages for each individual. [COLOR IS REQUIRED]

**Subject Scaling Factor**

After identifying the patterns of brain activity, the subject scaling factor (SSF) is the expression of the pattern for each contrast image or level of task demand. The following analyses are for SSFs derived from participant data after removing group effects. To summarize, the main effects of load and sex were significant when predicting the group-derived expressions. The interactions between load and sex and the main effect of load were significant within the participant-level data. These results are detailed below.

**Group level SSF**

Predicting expression of the group derived pattern by each individual, the main effect of load was significant ($F(4, 96) = 39.783$, $p < 0.001$) as was the main effect of sex ($F(1,23) = 5.130$, $p = 0.033$), see Figure 7. The interaction between load and sex was not significant ($F(4, 96) = 0.580$, $p = 0.678$). The main effect of cognitive capacity was not significant ($F(1,23) = 4.216$, $p = 0.0516$). Repeated differences between levels of task load demonstrate significant differences in expression between loads 1 to 2 ($t(96) = -3.746$, $p = 0.0003$), and loads 2 to 3 ($t(96) = -5.166$, $p < 0.0001$). The remaining differences were not significantly different: loads 3 to 4 ($t(96) = -1.186$, $p = 0.238$) and loads 4 to 5 ($t(96) = 0.287$, $p = 0.775$). The main effect of sex was driven by greater expression by the males than females ($t(23) = 2.265$, $p = 0.033$). The random component of the model (participant, intercept) was significant (ICC = 0.548, $X^2(1) = 47.354$, $p < 0.0001$).

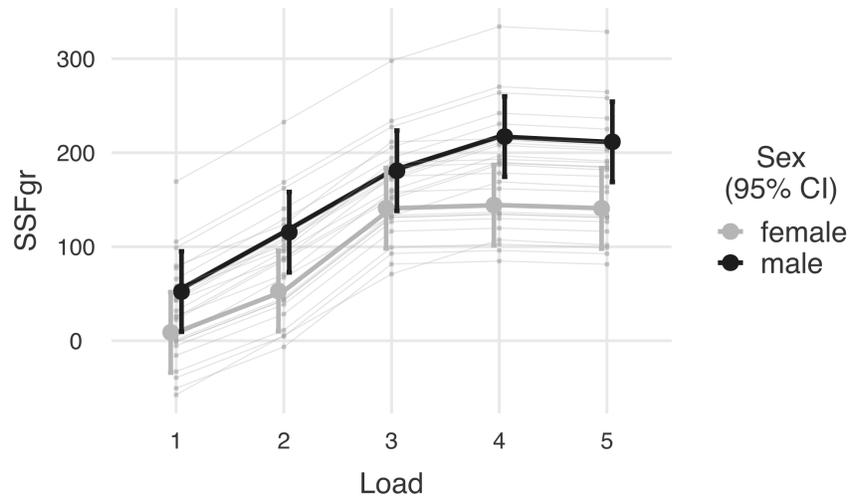

Figure 7. Cognitive load by expression of subject scaling factors for the group derived principal component. Results demonstrate the significant effects of load and sex on expression. Each light gray line represents a different individual as modeled with the linear mixed-level modeling. [COLOR IS NOT REQUIRED]

**Participant level SSF after removing group-level effects**

Predicting expression of the individually derived patterns after removing group effects, the interaction between load and sex was significant ($F_{(4, 96)}$ = 4.880, p = 0.0013) as was the main effect of load ($F_{(4, 96)}$ = 69.029, p < 0.001), see Figure 8. The main effect of cognitive capacity was not significant ($F_{(1,23)}$ = 1.498, p = 0.233), nor the main effect of sex ($F_{(1, 23)}$ = 0.114, p = 0.739). Repeated differences between levels of task load crossed with sex demonstrate significant differences in expression between loads 3 to 4 (t(96) = -3.132, p = 0.0023). The remaining differences were not significantly different: loads 1 to 2 (t(96) = 1.683, p = 0.100), loads 2 to 3 (t(96) = 1.175, p = 0.243) and loads 4 to 5 (t(96) = -0.671, p = 0.504). The random component of the model (participant, intercept) was significant (ICC = 0.240, $X^2$(1) = 9.895, p = 0.0017).

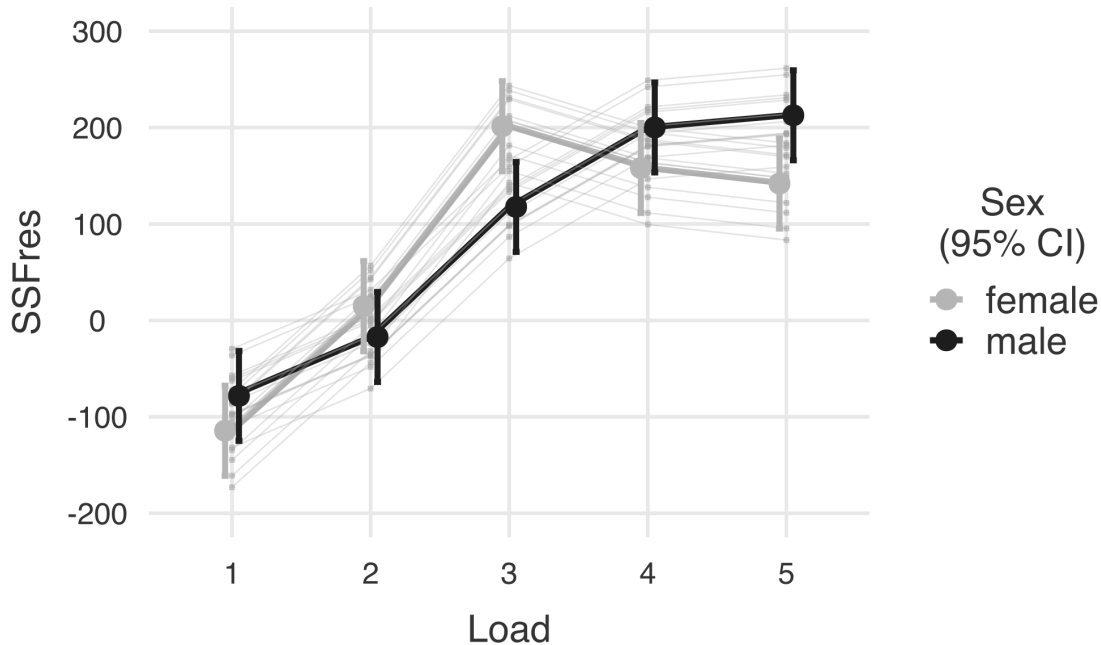

Figure 8. Cognitive load by expression of subject scaling factors for the participant level derived principal component. Results demonstrate the significant interaction of load and sex on expression. Each light gray line represents a different individual as modeled with the linear mixed-level modeling. [COLOR IS NOT REQUIRED]

**Predicting Performance with SSFs**

The behavioral data analyses presented above are now presented with the inclusion of the expression of the group derived and participant-specific patterns of brain activity to test for brain-behavior relationships. Results demonstrate that the brain measures are unrelated to the measure of task accuracy; however, response time was significantly predicted by expression of the participant-specific patterns of brain activity and not by expression of the group-derived pattern.

**Brain Activity and Accuracy**

Predicting accuracy, the main effect of load was significant (F(4, 102.602) = 7.714, p < 0.001). No interaction effects were significant: load by sex (F(4, 91.878) = 0.674, p = 0.611), load by $SSF_{group}$ (F(4, 99.979) = 0.300, p = 0.877), and load by $SSF_{participant}$ (F(4, 105.259) = 0.941, p = 0.443). The remaining main effects were also not significant: cognitive capacity (F(1,26) = 1.138, p = 0.296), sex (F(1,27.142) = 0.014, p = 0.907), $SSF_{group}$ (F(1, 58.981) = 0.191, p = 0.664), and $SSF_{participant}$ (F(1, 95.307) = 0.450, p = 0.504). Repeated differences between levels of task load demonstrate significant differences in accuracy between loads 3 and 4 (t(92.887) = 2.537, p = 0.0129). The remaining differences were not significant: loads 1 and 2 (t(107.368) = -0.253, p = 0.801), loads 2 and 3 (t(107.265) = 1.249, p = 0.214), and loads 4 and 5 (t(97.141) = 1.351, p =

0.180). The random component of the model (participant, intercept) was not significant (ICC = 0.025, $X^2(1) = 0.125$ p = 0.724).

**Brain Activity and Response Time**

Predicting response, the interactions between load and $SSF_{participant}$ was significant (F(4, 88.763) = 3.197, p = 0.017). The main effects of load (F(4, 89.523) = 5.320, p < 0.001) and cognitive capacity (F(1, 23.861) = 7.258, p = 0.013) were both significant. The main effect of cognitive capacity was driven by greater overall expression of brain activity for those with higher cognitive capacity. The remaining interactions were not significant: load by sex (F(4, 86.734) = 0.341, p 0.850) and load by $SSF_{group}$ (F(4, 87.819) = 0.956, p = 0.436). The other main effects were also non significant: sex (F(1, 24.315) = 2.841, p = 0.105), $SSF_{group}$ (F(1, 104.606) = 0.019, p = 0.889), and $SSF_{participant}$ (F(1, 95.237) = 0.560, p = 0.441). Repeated differences between levels of task load crossed with SSFparticipant demonstrate significant difference in response times between loads 3 and 4 reflecting those with the lowest expression of brain activity had large increases in brain activity and response times and those with high overall levels of brain activity were minimally differences between the load levels, see Figure 9, (t(87.192) = 3.284, p = 0.001). The remaining differences were not significant: loads 1 and 2 (t(89.409) = -0.198, p = 0.844), loads 2 and 3 (t(88.333) = -0.033, p = 0.973), and loads 4 and 5 (t(89.264) = -1.757, p = 0.082). The random component of the model (participant, intercept) was significant (ICC = 0.692, $X^2(1) = 70.304$ p < 0.001).

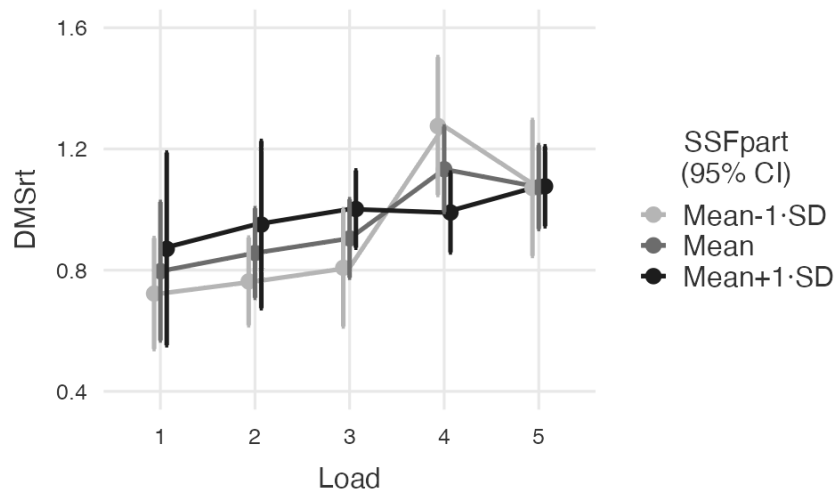

Figure 9. Cognitive load by response time in seconds split by level of expression of the SSF derived from each participant ($SSF_{part}$). Results demonstrate a significant interaction between load and pattern expression. [COLOR IS NOT REQUIRED]

## Discussion

This study identified distinct patterns of load-related brain activity common to the group and specific to each participant. The participant-specific patterns of brain activity were all expressed in similar load-related manners despite each involving different brain regions. Therefore, regardless of these regional differences, the individuals all used their own specific patterns of brain activity in a similar load-related manner. Brain activity increased as task demands increased in a load-related manner and either plateaued at an individual's cognitive capacity or continued to increase. Expression of the group pattern demonstrated a similar relationship and plateau with increasing load. In addition, expression of the participant-specific patterns of brain activity were better predictors of response time task performance than the expression of the group-derived pattern.

Despite qualitative similarities in the load-related expression of the group and participant-specific brain activity patterns, there are several significant quantitative differences. The expression of the group derived pattern differed by sex, with the male participants having consistently greater expression at all levels of task load. This sex finding contradicts a recent review of sexual dimorphisms in brain imaging research which found limited sex differences (Eliot et al., 2021). The current results suggest that sex differences exist; however, the differences become evident when looking across a wide range of task demands and are missed when comparing brain activity at one or two different load levels. It is also possible that the current experimental design of comparing brain activity at matched levels of difficulty instead of task load made sex differences evident.

Results demonstrated a trend-level effect between greater expression of the group-derived pattern across load levels and greater cognitive capacity. It is doubtful that this group-level derived brain-behavior effect would be significant with larger samples. This conclusion is due to recent findings that larger samples decreased the strength of brain-behavior relationships because of the multitude of ways individuals utilize their brains to perform cognitive tasks (Grady et al., 2021). A large amount of variance not accounted for in the model provides additional support for this argument.

After removing the group effect from the individual-level data, the participant-level brain maps were derived. Using the expression of these patterns, the same model used with the group data now left less than one-quarter of the variance between participants unaccounted for. The group pattern's variance accounted for from each participant's load level patterns also shows a wide range of values. The range between participants was between 1.5 and 57.5%. Therefore, the group pattern is not a good representative of the wide range of individual-level differences between people, thus supporting the claims by Grady et al. (2021).

The expression of participant-specific patterns of brain activity demonstrates significant interactions between load and sex and between load and cognitive capacity. The females demonstrated a peak in their expression scores at the third load level and then a decline for load levels at and above their respective cognitive capacities. The males demonstrate a continual increase in expression across all load levels. Greater expression at a higher load level is also related to greater cognitive capacity. Although the three-way interaction between load, sex, and cognitive capacity is not significant, marginal means demonstrate this load by cognitive capacity effect is solely in the females. Therefore, while the males benefit from increased expression of the group-derived pattern, there is support that females benefit from the expression of their individualized patterns. Future work will need to confirm this.

The participant-specific brain activity patterns were utilized similarly across load levels despite the limited overlap in the brain regions involved. At most, seven people commonly utilized any single voxel. The lack of overlap may not be surprising due to removing the central tendency from the data, which may have captured all common areas of brain activity. However, it is surprising to see the significant load effect in both the expression of the group and participant-derived patterns. Expression of the participant-specific patterns also better predicted response time task performance than the expression of the group pattern.

Response time performance increased as a function of load level. Accuracy performance decreased. The lack of relationship between accuracy and brain activity is by design. The task demands were chosen such that accuracy would be approximately 80% at load level four and lower above that. Therefore, there is minimal between-participant variance in accuracy across load levels. Load-related increases in expression of the participant level pattern, but not the expression of the group pattern, significantly predicted response time when including both in the model.  Larger cognitive capacity also predicted greater response time. This observation is due to the nature of the experimental design itself. Each participant's specific load levels were chosen based on their cognitive capacity. Therefore, someone with a higher cognitive capacity received task demands with higher load levels. Therefore, this finding explains the positive relationship between response time and load level (S. Sternberg, 1966; Saul Sternberg, 2016).

The observed load-related responses of brain activity fit into current theories of cognitive aging. Neural capacity describes brain activity reaching a maximum level as task demands increase (Stern, 2009; Stern et al., 2005). When brain activity decreases with increasing task demands after reaching a neural capacity, the activity is described as Compensation-Related Utilization of Neural Circuits Hypothesis (CRUNCH) (Reuter-Lorenz & Lustig, 2005). In previous work with the CRUNCH thesis, differences in brain activity as a function of cognitive load were largely attributed to individual differences in working memory span (Schneider-Garces et al., 2010). These authors demonstrated this through post-hoc adjustment of their imaging data by measures of task accuracy. This approach was unnecessary in the current work since each participant's brain imaging data was collected across subjective levels of task demand, negating the need for data adjustments.

The current results, however, did not support CRUNCH for two possible reasons. One is that the experiment used task demands controlling for perceived difficulty. Therefore, the greatest level of task demand was only one load item above someone's cognitive capacity. Observations of declines in brain activity may require task demands well above someone's cognitive capacity. It is also possible that CRUNCH occurs primarily in older adults and not younger adults like the current experiment. It will be interesting to explore the role of the higher-order principal components in young and old adults to identify if patterns of brain activity remain stable across load levels in both young and old adults. The current work assumes they do.

These analyses retained only the first of four PCs from each individual. These primary PCs accounted for between 39 and 77% of the variance in each participant. There may be secondary functional processes occurring as task demands increase which the other PCs would capture. Exploring higher PCs has been done with group-level analyses (Stern et al., 2008; Zarahn et al., 2006). Within participants, some may employ alternate strategies (Miller et al., 2012) at different load levels, thereby utilizing different patterns of brain activity depending on the task demands. Future work will explore this avenue of thought. However, unlike Miller et al. (2012), the current methods can use data-driven analyses to identify alternate strategies based on the brain imaging data and not post-hoc behavioral assessments.

This work is novel in that it uses a methodology that relies on an individual's pattern of brain activity at their subjective level of task demands. It does not simply explore whether an individual's variance around a group central tendency at common load levels is predictive of their behavior. However, despite individual differences in patterns of brain activity, there are strong commonalities in the utilization of respective patterns. These methods meet the recent suggestions that approaches are needed to understand between-participant variance in brain imaging data (Lebreton et al., 2019). Furthermore, these methods provide avenues for investigating individual differences in using neural resources to meet task demands (Seghier & Price, 2018). Finally, the current work provides a means of addressing the concerning observation that unaccounted for individual-level variance in brain activity weakens group analyses of brain-behavior relationships (Grady et al., 2021).

Many questions are addressable with these methods beyond the capabilities of group analyses. One is the exploration of different cognitive strategies. As mentioned above, the current work only investigated the first PC. It is also noted that this first PC accounted for a wide range of variance. Large expressions of higher-order PCs may be indicative of an individual utilizing multiple distinct patterns of brain activity. These may reflect different cognitive strategies across different levels of task demands. A second question would be an exploration into physiological differences underlying the individuality in patterns of brain activity. Exploring physiological differences could be done using analyses that fuse structural and functional brain measures. Such approaches include multivariates fusion analysis (Sui et al., 2014) or voxel-wise serial univariate mediation analyses (Steffener et al., 2016). Work with older adults experiencing age-related neural changes will explore this possibility. It is also possible that analyses including genetic, developmental, or lifetime exposures (Steffener & Stern, 2012; Stern, 2002) would shed light on the individual differences in patterns of brain activity. Future work with larger samples will explore how load-related differences in pattern expression differ as a function of individual differences, allowing for the development of brain activity profiles.

There are several limitations in the current work and findings. There is a relatively small sample size of twenty-six. However, the results support the recent demonstration that larger samples are not always better (Grady et al., 2021). In addition, the mixed-level statistical modeling used does not provide standardized effect sizes. Unfortunately, due to how these models partition variance (Rights & Sterba, 2019), there is no agreed-upon way for calculating standardized effect sizes. We instead now report the unstandardized effect sizes of the fixed effect in line with general recommendations (Pek & Flora, 2018). Furthermore, this study did not account for differences in the women's menstrual cycle phase when testing occurred (Dubol et al., 2021).

Despite these limitations, there are multiple strengths of the current work. This study used individualized patterns of brain activity derived over a wide range of cognitive demands. Pre-assessment of each individual's working memory cognitive capacity allowed task demand delivery in the MRI at the same perceived difficulty level for all individuals. Using linear mixed statistical models is also superior at controlling type I errors than alternative repeated measures ANOVA models. Therefore, results from mixed models have a greater likelihood of generalizing to new data sets (Barr et al., 2013; Judd et al., 2012).

## Conclusion

Participants in this sample demonstrated a wide range of distinct patterns of brain activity when performing the same task at matched levels of perceived difficulty. Despite spatial differences in patterns of brain activity, there were substantial similarities in how the expression, or usage, of these patterns changed as task demands increased. The expression of the participant-specific patterns of brain activity were also better predictors of task performance than the expression of a group-derived pattern. Furthermore, the expression of the group and participant-specific patterns differed between the sexes.

It may be time to reassess some of the main assumptions implicit in the field of neuroimaging. The use of central tendencies has a goal of identifying commonality in the brain regions involved in a task and assumes each person uses the same brain regions in the same manner to perform a task. The current results demonstrate similar task-related responses of brain activity but in a wide range of regions. The use of the same task demands for all individuals also assumes a similar perception of the task across all participants. However, as the field further explores individual differences in cognitive strategy and physiological variations in brain activity, this assumption also needs reassessment. Finally, sexual dimorphism in brain activity may only be evident when individual differences in patterns of brain activity are incorporated into analyses.

## Acknowledgements

Thank you to Rahim Ismaili at the BIC for MRI scanning.

## Funding

This work was supported by University of Ottawa funds to JS.